\documentclass[aps,prl,preprint,groupedaddress]{revtex4-1}
\usepackage{amsmath}
\usepackage{epsfig}
\usepackage{longtable}
\usepackage{float}
\newfont{\logo}{logo10}

\newcommand{\bea}{\begin{eqnarray}}
\newcommand{\eea}{\end{eqnarray}}
\newcommand{\bes}{\begin{subequations}}
\newcommand{\ees}{\end{subequations}}

\begin{document}

%\preprint{}

\title{Harnessing energy sharing collisions of Manakov solitons to implement universal NOR and OR logic gates}


\author{M. Vijayajayanthi\footnote{ e-mail: vijayajayanthi.cnld@gmail.com}}
%\thanks{}
\affiliation{Department of Physics,
 \\B. S. Abdur Rahman Crescent Institute of Science and Technology,
 \\ Chennai--600 048, India}
\author{T. Kanna\footnote{Corresponding author e-mail: kanna$\_$phy@bhc.edu.in}}
\affiliation{PG and Research Department of Physics,
 \\
Bishop Heber College,Tiruchirapalli--620 017, India}
\author{K. Murali\footnote{ e-mail: kmurali@annauniv.edu}}
\affiliation{Department of Physics, Anna University, Chennai--600 025, India}
\author{M. Lakshmanan\footnote{ e-mail: lakshman@cnld.bdu.ac.in}}
\affiliation{Centre for Nonlinear Dynamics, School of Physics, Bharathidasan University, Tiruchirapalli--620 024, India\\}

%\date{\today}

\begin{abstract}
The energy sharing collision of bright optical solitons in the Manakov system, governing  pulse propagation in high birefringent fiber, is employed theoretically to realize optical logic gates.  Especially, for the first time, we successfully construct (theoretically) the universal NOR gate and the OR gate from the energy sharing collisions of just four bright solitons which can be well described by the exact bright four-soliton solution of the Manakov system.  This construction procedure has the important merits like realizing the two input gates with a minimal number of soliton collisions and possibilities of multi-state logic.  The recent experiments on Manakov solitons suggest the possibility of implementation of this theoretical construction of such gates and ultimately an all optical computer.
\end{abstract}

%\keywords{}

\maketitle

\section{}
Solitons are fascinating mathematical objects which arise as solutions of certain integrable nonlinear evolution equations \cite{ablowitz}. Their remarkable collision behaviour and other dynamical properties have led them to find applications in several fields ranging from water waves \cite{whitham}, plasma physics \cite{infeld}, nonlinear optics \cite{kivshar} to Bose-Einstein condensates (BEC) \cite{kevrekidis}. Especially in optics, they result due to a delicate balance of natural dispersive spreading of the wave by an inherent nonlinearity, namely Kerr nonlinearity.  Solitons in single mode optical fibres were first predicted by Hasegawa and Tappert \cite{hase1} and experimentally confirmed subsequently by Mollenauer \textit{et al}. \cite{molle}.  Following this, there has been a great surge of studies on optical solitons.  One important aspect of such a study which is recently receiving attention is multicomponent  solitons (MSs)\cite{kevrekidis1}.  Some examples of multicomponent solitons are partially coherent solitons \cite{akhmediev}, soliton complexes \cite{soto}, multi-mode pulses in optical fibers \cite{mecozzi}, symbiotic solitons \cite{abdullaev}, spinor solitons \cite{ieda}, etc.

The work horse for such enormous studies on MS is the celebrated Manakov model \cite{man}, which is also known to be integrable \cite{radhakrishnan}. In optics, the Manakov system governs the propagation of a  pair of orthogonally polarized high intense optical pulses in single mode optical fibers.   The interplay between the dispersion and the self phase modulation (phase shift in a given mode depending upon its own intensity) as well as the cross phase modulation (phase shift due to the intensity of the co-propagating mode) effects  result in optical solitons.  Recent developments in photonic crystal fibers have lead to significant developments in the experimental observation of such optical solitons \cite{philbin}.

The incoherently (intensity dependent nonlinearity) coupled nonlinear Schr\"odinger system describing the propagation of two orthogonally polarized high intense optical pulses in an elliptically birefringent fiber with high birefringence can be casted as \cite{agarwal1},
\bes
\bea
i(Q_{1\zeta}+\beta_{1x}Q_{1\tau})-\frac{\beta_{2}}{2}Q_{1\tau\tau}+\gamma(|Q_1|^2+B|Q_2|^2)Q_1=0,\\
i(Q_{2\zeta}+\beta_{1y}Q_{2\tau})-\frac{\beta_{2}}{2}Q_{2\tau\tau}+\gamma(|Q_2|^2+B|Q_1|^2)Q_2=0,
\eea
 \label{1}
\ees
where $\zeta$ and $\tau$ are respectively propagation direction and time, $Q_j$'s, $j=1,2,$ are complex slowly varying amplitudes, $\beta_{1x}$ and $\beta_{1y}$ are inverse of group velocities of two modes, $\beta_{2}$ represents group velocity dispersion (GVD) and the effective Kerr nonlinearity coefficient $\gamma$ is defined as $\frac{8 n_2 \omega_0}{9 c A_{eff}}$, where $n_2$ is the nonlinear index coefficient, $\omega_0$ is the carrier frequency and $A_{eff}$  is the effective core area.  Here $\gamma$ and $\beta_2$ are same for both the pulses as they are at the same wavelength.
The cross phase modulation coupling parameter $B=\frac{2+2\sin^2\theta}{2+cos^2\theta}$, where $\theta$ is the ellipticity angle which can vary between 0 and $\pi/2$.
%Here propagation distance and time are measured in kilometer ($km$) and in picosecond ($ps$).
%By introducing the transformations $Q_j=q_j\; e^{i(\frac{\sigma^2 z}{2}+(-1)^j\sigma t)}$ and $z'=2z$ and again redefining $z'$ as $z$, we  arrive at the standard Manakov model
% \cite{man},
%\bea
%i q_{j,z}-\beta_{2} q_{j,tt}+2 \gamma\sum_{l=1}^2 |q_l|^2\;q_j=0,\;\;j=1,2. \label{manakov1}
%\eea
%Note that by considering soliton units Eq. (\ref{manakov1}) can be expressed in dimensionless form \cite{agarwal1} where $\beta_2$ and $\gamma $ can be absorbed appropriately. However, here we consider Eq. (\ref{manakov1}) as it is with $\beta_2 <0$, corresponding to anomalous dispersion region and for convenience we assume $|\beta_2|$ to be unity.
For lossless fibres, after suitable transformations, the above equation (\ref{1})  can be expressed in the following dimensionless form using soliton units \cite{agarwal1},
\bes
\bea
i Q_{1z}-\frac{sgn(\beta_2)}{2} Q_{1tt}+\mu^2(|Q_1|^2+B|Q_2|^2)Q_1=0,\\
i Q_{2z}-\frac{sgn(\beta_2)}{2} Q_{2tt}+\mu^2(|Q_2|^2+B|Q_1|^2)Q_2=0,
\eea
\ees
where the dimensionless length and retarded time are defined as $z=\frac{\zeta}{L_D}$, $t=\frac{T}{T_0}=(\tau-\tilde\beta_1\zeta)$ in which the dispersion length $L_D=\frac{T_0^2}{|\beta_2|}$, nonlinear length $L_{NL}=\frac{1}{\gamma P_0}$ and $\tilde\beta_1=\frac{1}{2}({\beta_{1x}+\beta_{1y}})$ with $T_0$ and $P_0$ being the initial width and peak power, $\mu^2=\frac{\gamma P_0 T_0^2}{|\beta_2|}$.
In the anomalous (normal) dispersion regime, $\beta_{2}<0~(>0)$, where the high (low) frequency pulses travel faster than the low (high) frequency pulses, the above equation is referred as focusing (defocusing) coupled nonlinear Schr\"odinger (CNLS) equation and the fibre supports bright (dark and dark-bright) solitons.  These are consequences of the polarization modulation instability \cite{baronio}.  For the polarizing angle $\theta=35^\circ$ (for which $B=1$) in the anomalous dispersion regime, with trivial transformations $z^{\prime}=\frac{z}{2}$, $q_j=\mu Q_j, j=1,2,$ and dropping the prime, we get the standard Manakov model in normalized form as  \cite{man},
\bes
\bea
i q_{1z}+ q_{1tt}+2(|q_1|^2+|q_2|^2)q_1=0,\\
i q_{2z}+ q_{2tt}+2(|q_1|^2+|q_2|^2)q_2=0.
\eea
\label{manakov1}
\ees
The Manakov system (\ref{manakov1}) has been extensively studied (for details see \cite{man,hie,kannaprl,kannapre,dinda,kanapramana,epj} and references therein). The striking feature of this Manakov system is the fascinating  energy sharing collision of bright solitons as a consequence of change in the polarization vector during collision. In such energy sharing collision, the intensity of soliton in a given component can be enhanced (suppressed) while the other soliton experiences an opposite effect. Contrary to this the solitons in the second component display a reverse scenario thereby preserving the total intensity as well as the intensity of the individual component \cite{hie,kannaprl,kannapre,dinda,kanapramana,epj}.  Following this, multisoliton interactions in various multicomponent CNLS type systems including the Manakov system have been studied in Refs. \cite{kannapre,kanapramana,epj}.  These Manakov bright solitons have been experimentally realized in $Al_x Ga_{1-x} As$ waveguides \cite{expt} and their energy sharing collision has also been experimentally demonstrated \cite{ana}.  Recently, optical dark rogue waves have also been experimentally observed in the Manakov model with defocusing nonlinearity \cite{millot}. Thus the Manakov solitons are suitable candidates for experimental realization and for further technological applications.  Also, in Ref. \cite{josa} it has been   clearly demonstrated that this energy sharing property of the Manakov solitons is preserved in the presence of fibre losses and is robust against strong environmental perturbations. This shows that the present construction procedure will hold good even in the presence of losses.

From a technological point of view, in the era of digital electronics, integrated circuits are usually made up of universal NOR and  NAND gates which are considered to be the basic modules of all other logic gates. However, modern day computers have their own demerits such as heat dissipation, processing speed, space, and speed of transmission, etc. \cite{book,murali,sapin,coupler}.  To overcome these difficulties, many researchers proposed that the light field, especially solitons, can act as the carrier of information instead of electrons which we employ in our present day computers \cite{ada_BZe,jaku,steig,rand,pramana,miller}. As an important advancement in optical computing, in Ref. \cite{miller}, the criteria required for practical optical logic (POL) are nicely discussed in detail. In Ref. \cite{jaku} energy sharing collision of Manakov solitons was profitably used for performing nontrivial information transformation. Especially, sequences of solitons operating on other sequences of solitons effect logic operations. Later, Steiglitz \cite{steig} theoretically constructed various gates such as the COPY, FANOUT, Z  and Y converters and combined them to realize the NAND gate. In Ref. \cite{steig}, through numerical simulation separate sequential collisions of non-interacting data and operator solitons were considered where the operator soliton always remains unaffected.

The main task of this proposal is to use the pair-wise energy sharing collisions of bright Manakov solitons as such without imposing any constraints on the colliding solitons for realizing the universal logic gate. In such a collision process, all solitons undergo energy sharing collisions and every soliton interacts with every other solitons which are involved in the collision process.  To be specific, we employ just a four bright soliton collision process, in which each soliton undergoes three pair-wise energy sharing interactions.   We have recently constructed single input logic gates such as COPY gate, NOT gate and ONE gate using the energy sharing collisions of three bright optical solitons associated with the three soliton solution of the Manakov system \cite{one_gate}.  However, it is not a straightforward task to extend this study for the realization of two input gates due to the cumbersome form of the N-soliton solution of the Manakov system with $N>3$. Also there is no clue about the number of solitons required.  Here, we succeed by a careful albeit tedious asymptotic analysis to identify that collision of four solitons is sufficient to construct the universal NOR gate in a more practical physical situation.  Indeed, the computation occurs by the pair-wise energy sharing collisions of solitons, where each soliton bears a finite state value before collision, and state transformations occur at the time of collisions between solitons. The novelty of the present work is to realize the universal NOR gate in a theoretical sense by utilizing the energy sharing collisions \cite{hie,kannaprl,kannapre} of only a minimal number of four solitons arising in a high birefringent telecommunication fibre. Other physical systems where the energy sharing collision can be observed and hence suitable for computing are multi-species BEC \cite{kevrekidis}, photorefractive materials \cite{photoref} and left handed materials \cite{lhm}, etc.

Here, we consider the interaction of the four bright solitons in the Manakov system, described by four soliton solution given in the supplemental material \cite{supplement}. As pointed out earlier, during the energy sharing collision, the Manakov solitons experience a change in their states (polarizations) due to the enhancement or suppression of intensity which is the desirable property for performing computation. Also, it is sufficient to examine these states well before (i.e., at the input) and well after (output) collisions. The key idea is to define the asymptotic states of the $j^{th}$ soliton as
$\rho^{j\pm}=\frac{q_1^j(z\rightarrow\pm\infty)}{q_2^j(z\rightarrow\pm\infty)}=\frac{A_1^{j\pm}}{A_2^{j\pm}},$ where $A_{1,2}^{j\pm}$ are the polarization components (1,2) of the $j^{th}$ soliton. Here suffices denote the components, $+(-)$ designates the state after (before) collision and superscript $j$ represents the soliton number. The logic gates deal with binary logic, either $1$ or $0$. We define such $1(0)$ state if the intensity $|\rho^{j_{\pm}}|^{2}$ of the state vector exceeds (falls below) a particular reference value. This clearly shows that our construction procedure avoids the critical biasing, a property for POL mentioned in Ref. \cite{miller}. Additionally, as we are dealing with states defined by the ratio of intensities, the important differential signalling criteria required for POL \cite{miller} is also naturally satisfied.

Also, we may note that the Manakov solitons undergo pair-wise collisions. In our construction procedure, we are going to consider four solitons which interact in a pair-wise manner and we denote the input solitons as unprimed solitons $S_{j}, j=1,2,3,4$. We will refer the solitons emerging after the first, second and third pair-wise collisions as primed, double primed and triple primed solitons respectively. In fact $S_{j}^{'''}$ represent the output solitons.  A schematic diagram of this collision process is shown in the supplemental material \cite{supplement}. The intensities of the four colliding solitons at the input and at the output are calculated analytically from a systematic but rather lengthy asymptotic analysis, presented in the supplemental material \cite{supplement}. Here, we assume that the soliton parameters $k_{jR}>0$; $k_{1I}>k_{2I}>k_{3I}>k_{4I}$ ($k_{jR}=$Re$(k_{j})$, $k_{jI}=$Im$(k_{j})$), where the suffices R and I denote the real and imaginary parts.

For constructing the universal NOR gate, the inputs are fed into the solitons $S_1$ and $S_2$ before interaction, and the output is taken up from the soliton  $S_4$ after interaction. Thus the input and output solitons are treated separately which will prevent the input pulses to be reflected back into the output pulse, a criteria referred as input/output isolation necessary for POL. The explicit forms of the states of the solitons $S_1$ and $S_2$ before interaction are
\bea
\rho^{1-}&=&\frac{\alpha_1^{(1)}}{\alpha_1^{(2)} },\\
\rho^{2-}&=&\frac{A_1^{2-}}{A_2^{2-}}=\frac{N_1^{2-}}{N_2^{2-}}=\frac{\alpha_1^{(1)} \kappa_{21}-\alpha_2^{(1)} \kappa_{11}}{\alpha_1^{(2)} \kappa_{21}-\alpha_2^{(2)} \kappa_{11}}.
\eea
Similarly, the state of the soliton $S_4$ after collision is given by
$\rho^{4+}=\frac{\alpha_4^{(1)}}{\alpha_4^{(2)} }$.
In the above equations,  $\alpha_{l}^{(m)},l=1,2,4, m=1,2,$ represent the polarization parameters of solitons $S_1$, $S_2$ and $S_4$ and they can take any arbitrary complex value.  Similarly, the other quantities $\kappa_{11}, \kappa_{21}$, and $\kappa_{12}$ are defined by the soliton parameters $\alpha$'s and $k$'s.  Though $S_3$ does not explicitly appear in the above expression, it indirectly influences the energy sharing collisions. The ratio of intensities of the solitons $S_1$, and $S_2$ before interaction as well as the soliton $S_4$ after interaction can be obtained by taking the absolute squares of these complex states and they are given by $|\rho^{1-}|^2$, $|\rho^{2-}|^2$ and  $|\rho^{4+}|^2$, respectively.  Hence, one can measure the ratio of the intensities of the input/output solitons analytically from the asymptotic analysis. Then as mentioned before if the ratio of intensities of a given  soliton $S_j$ is greater (less) than some specific threshold value, say 1, before interaction then we denote the  input state of $S_j$ as $``1 (0)"$ state.  Thus $1$($0$) state of soliton $S_j$ corresponds to $|\rho^{j-}|^2>1 (<1)$. To achieve the required output corresponding to the NOR gate from asymptotic analysis, we deduce the following condition on the soliton parameters:
\bea
\hspace{-0.3cm}\alpha_4^{(2)}=\left(\frac{\alpha_1^{(1)}}{\alpha_1^{(2)}}+\frac{\alpha_2^{(1)}\left((k_1-k_2)|\alpha_1^{(1)}|^2-(k_2+k_1^*)|
\alpha_1^{(2)}|^2\right)+(k_1+k_1^*)\alpha_1^{(1)}\alpha_1^{(2)*}\alpha_2^{(2)}}{(k_1+k_1^*)\alpha_1^{(1)*}\alpha_2^{(1)}\alpha_1^{(2)}-\alpha_2^{(2)}\left((k_2-k_1)
|\alpha_1^{(2)}|^2+(k_2+k_1^*)|\alpha_1^{(1)}|^2\right)}\right) \alpha_4^{(1)}.
\label{con2}
\eea
%where $R=\left(\frac{\alpha_1^{(1)}}{\alpha_1^{(2)}}+\frac{\alpha_2^{(1)}\left((k_1-k_2)|\alpha_1^{(1)}|^2-(k_2+k_1^*)|
%\alpha_1^{(2)}|^2\right)+(k_1+k_1^*)\alpha_1^{(1)}\alpha_1^{(2)*}\alpha_2^{(2)}}{(k_1+k_1^*)\alpha_1^{(1)*}\alpha_2^{(1)}\alpha_1^{(2)}-\alpha_2^{(2)}\left((k_2-k_1)
%|\alpha_1^{(2)}|^2+(k_2+k_1^*)|\alpha_1^{(1)}|^2\right)}\right)$.

\begin{figure}[!ht]
\begin{center}
\includegraphics[width=1\linewidth]{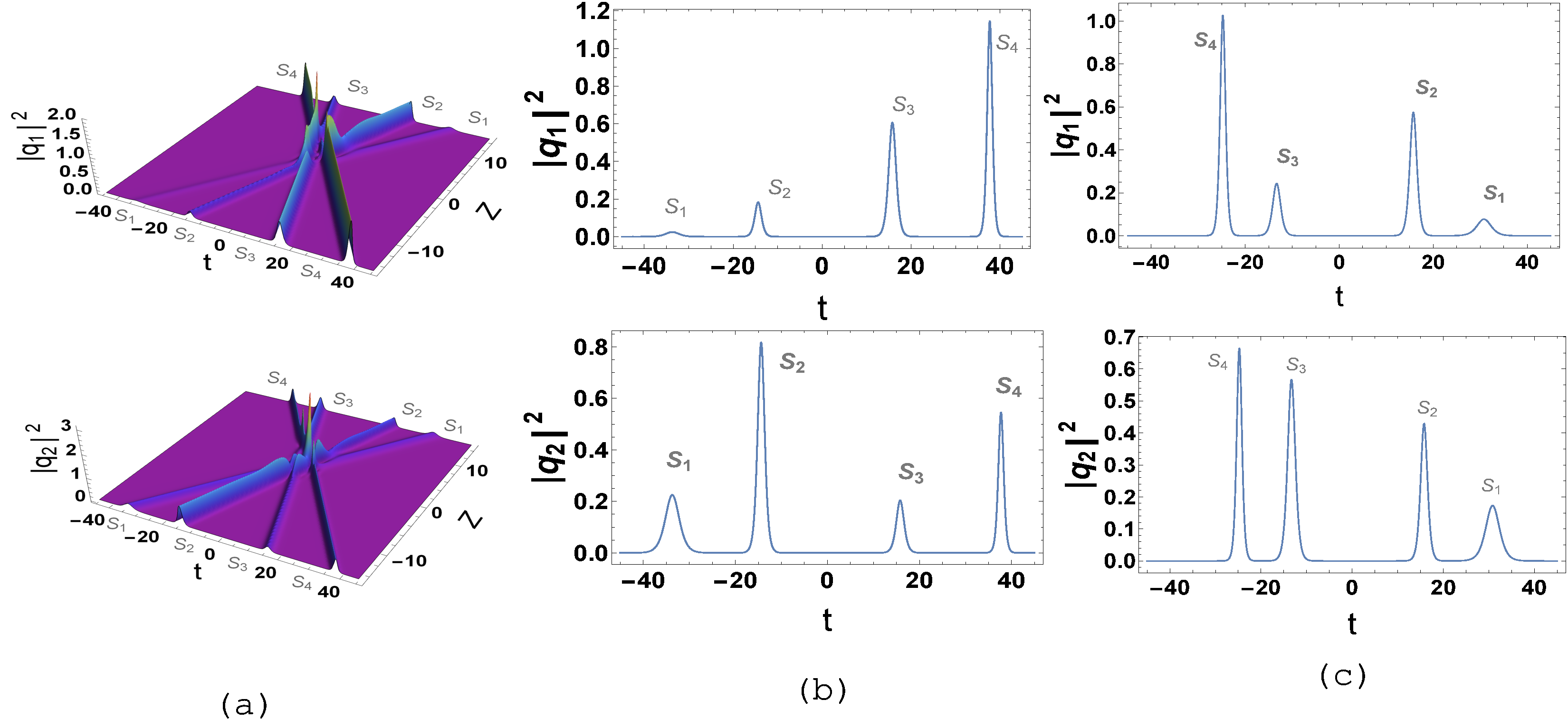}
\caption{NOR gate: The state of input solitons ($S_1$ and $S_2$) are $``0"$ and  $``0"$ and the state of output soliton ($S_4$) is $``1"$.  The first column (a) displays the mesh plots of the intensity profiles while the middle and last columns (b) and (c) depict the two dimensional plots of intensities at the input $(z=-15)$ and at the output $(z=15)$, respectively }\label{nor_00}
\end{center}
%\end{figure}
%\begin{figure}
\begin{center}
\includegraphics[width=1 \linewidth]{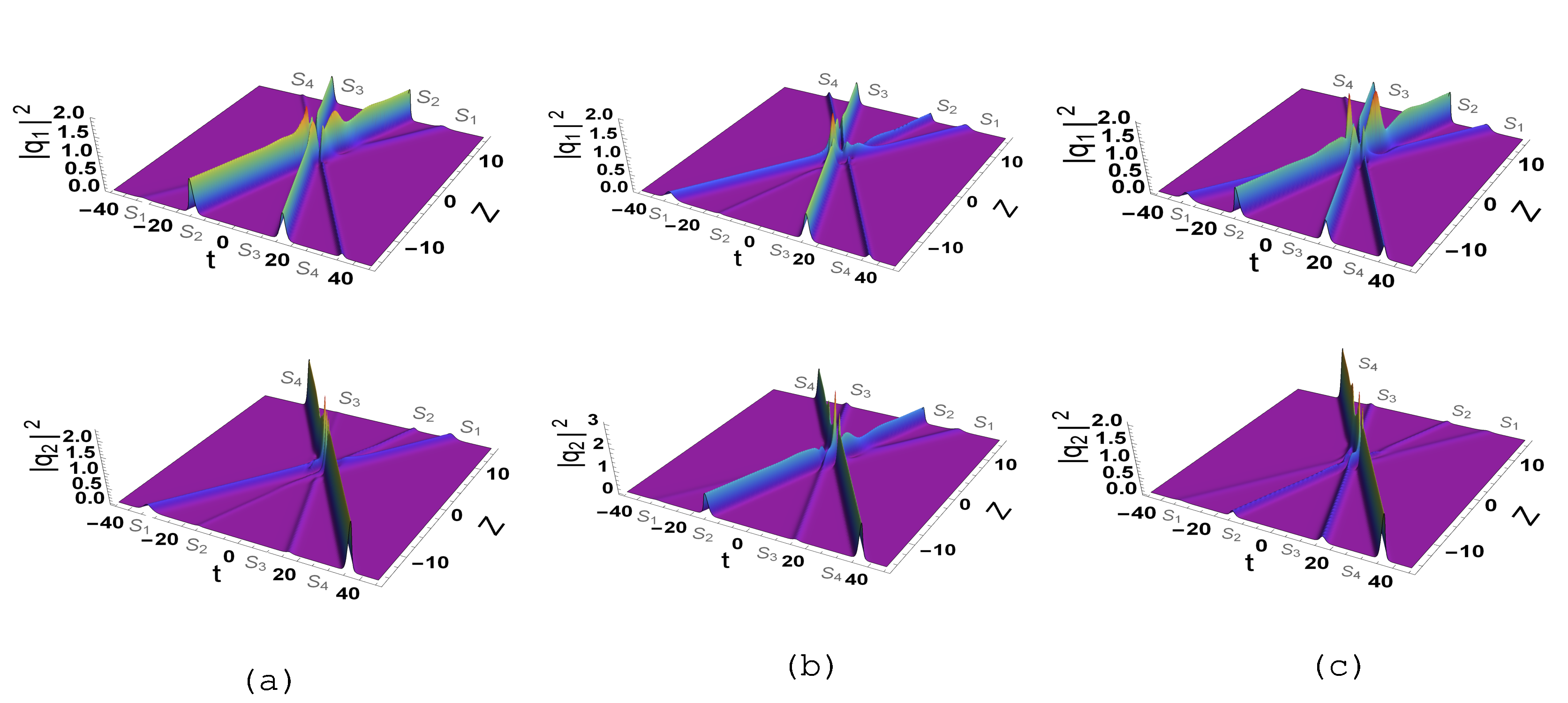}\label{nor_01_10_11}
\caption{NOR gate: (a) Input states  $``0"$ and  $``1"$, Output state  $``0"$; (b) Input states  $``1"$ and  $``0"$, Output state  $``0"$; (c) Input states  $``1"$ and  $``1"$, Output state  $``0"$  }
\end{center}
\end{figure}

The above relation (\ref{con2}) is obtained by imposing the condition, $\rho^{4+}=\left(\rho^{1-}+\rho^{2-}\right)^{-1}$ on the state vectors of the input solitons ($S_1$, $S_2$) and the output soliton ($S_4$) so that the Boolean algebra of the NOR gate is satisfied. Assigning $(0,0)$ input states to ($S_1$, $S_2$) by choosing $\alpha_1^{(1)}=2,  \alpha_1^{(2)}=6, \alpha_2^{(1)}=2, \alpha_2^{(2)}=5,$ we achieve the $``1"$ output state from soliton $S_4$ for the parameter choices $k_1=0.5+i, k_2=1+0.5 i, k_3=0.9-0.5 i, k_4=1.3-i, \alpha_3^{(1)}=3, \alpha_3^{(2)}=1, \alpha_4^{(1)}=0.001-0.002 i$ along with the condition (\ref{con2}), which is depicted in Fig \ref{nor_00}. This same choice is chosen for proving the rest of the three cases of the NOR gate whereas the input $\alpha_{i}^{(j)}$,$ i, j=1,2,$ parameters are varied. The quantities plotted in all the figures given in our work are in dimensionless form.  In Figure 1,  the first column (a) displays the mesh plots of the intensity profiles while the middle and last columns (b) and (c) depict the two dimensional plots of intensities at the input $(z=-15)$ and at the output $(z=15)$, respectively. Fig. 2(a) demonstrates that choosing ($0, 1$) input states to solitons ($S_1$, $S_2)$ for  $\alpha_1^{(1)}=2,  \alpha_1^{(2)}=6, \alpha_2^{(1)}=5, \alpha_2^{(2)}=2$, we get $``0"$ as the output state of $S_4$ after interaction.  Next,  $(1, 0)$ input states are assigned to solitons ($S_1$, $S_2$) and the $``0"$ output state is observed at the output of soliton $S_4$ for $\alpha_1^{(1)}=6,  \alpha_1^{(2)}=2, \alpha_2^{(1)}=2, \alpha_2^{(2)}=5$ which is shown in Fig. 2(b). Finally,  when $(1, 1)$ input states are given to solitons ($S_1$, $S_2$), the $``0"$ output state results for soliton $S_4$ for the parameter choices $\alpha_1^{(1)}=6,  \alpha_1^{(2)}=2, \alpha_2^{(1)}=5, \alpha_2^{(2)}=2$  as shown in the last column as Fig. 2(c).   The two dimensional plots of intensities corresponding to Fig. 2 are given in supplemental material for a better understanding.  The truth table and the corresponding intensity tables (calculated values of the ratios of intensities of solitons) are given in Tables I and II. It is clear that the input states ($``0"$ and $`1"$) are attained by adjusting properly the $\alpha_i^{(l)}, i,l=1,2,$ parameters as discussed above.  However, all the other soliton parameters except $\alpha_4^{(2)}$ can have arbitrary values while it is fixed by the condition (\ref{con2}) for the desired output that fulfills the truth table.

In a similar fashion, the OR gate can also be constructed from the condition
$\rho^{4+}=\rho^{1-}+\rho^{2-}$. The explicit form of the condition and the collision scenario leading to OR gate are provided in the supplemental material \cite{supplement}.
\begin{table}[!ht]
\begin{minipage}{.5\linewidth}
%\label{table1}
%{\scriptsize
\caption{Truth table of NOR gate}
%\centering
%\begin{ruledtabular}
\begin{tabular}
{|c|c|c|c|}
 % \cline{1-6}
\hline
 Input 1  & Input 2  & Output \\
 ($S_1$) & ($S_2$) & ($S_{4}'''$)  \\\hline
0& 0 & 1 \\ \hline
   0& 1& 0  \\ \hline
1& 0 & 0\\ \hline
1& 1 & 0  \\ \hline
\end{tabular}
%\label{table2}
%\end{table}
%\begin{table}
%\end{ruledtabular}
\end{minipage}%
\begin{minipage}{.5\linewidth}
\centering
%\scriptsize
\caption{Intensity table of NOR gate}
\begin{tabular}{|c|c|c|c|}
 % \cline{1-6}
\hline
Input 1 \;\; & Input 2 \;\;& Output \;\;\\
$|\rho^{1-}|^2$  (W) &$|\rho^{2-}|^2$  (W) &$|\rho^{4+}|^2$ (W)\\ \hline
 0.1& 0.2 & 2 \\ \hline
   0.1& 47& 0.02  \\ \hline
9& 0.02 & 0.1  \\ \hline
9& 5 & 0.03 \\ \hline
\end{tabular}
%\lable{table3}
\end{minipage}
\end{table}

Another advantage of this theoretical construction procedure is that the universal two input NOR gate can also be constructed by cascading the
output of the one input gate, a desirable property for POL.  To elucidate this, we point out that in a four soliton collision process, the first three solitons can be used to realize the one input copy gate \cite{one_gate} where the input at $S_2$ is copied at the output of $S_3$ (say $S_{3}''$).  Now, this $S_{3}''$ and the input at $S_1$ can act as the two inputs for the NOR gate while the output is taken away from the soliton $S_4$, say $S_{4}'''$ as usual.  This demonstrates the possibility of cascadability. Similarly, another important criteria of POL, namely Fan-out can also be achieved from our present theoretical construction.  Particularly, in a four soliton collision process, if the input state is fed into the soliton $S_1$ before collision, then it can be switched to the output of any other two solitons after collision, say solitons ($S_2$ and $S_4$) by appropriately imposing conditions on the soliton parameters.  This really implies the process of fanout which indicates the state of one soliton ($S_1$) before collision is used to drive as an input to at least two solitons ($S_2$ and $S_4$) after collision. The details of this will be presented elsewhere.

Here the optical pulses propagate in the form of solitons which are by nature well localized structures that can travel over long distances without alteration in shape. This special property of solitons can restore the
logic signal throughout its propagation in an optical fibre. Hence, our theoretical work completely satisfies all the criteria necessary for realizing POL.  This clearly demonstrates the strength and versatility of our theoretical construction of universal logic gate which will have significant impact in realizing optical logic.

In summary, we have demonstrated theoretically the construction of the universal gate, namely the NOR gate as well as the OR gate using the energy sharing collision of four bright solitons in a high birefringent fiber described by the famous Manakov system.  Here the computing is performed by analyzing the asymptotic state variations of the colliding solitons that follows from a detailed asymptotic analysis of the explicit four-soliton solution of the Manakov system. We have demonstrated systematically that by altering the polarization parameters of the input solitons $S_1$ and $S_2$, one can realize the favourable output from the soliton $S_4$ without adjusting any other parameters.  This implementation of universal NOR gate is quite interesting and will provide the gateway for the experimentalists to realize the optical logical gates including the universal gate.  We wish to remark that our theoretical construction of logic gates very well satisfies all the criteria required for POL recently discussed in Ref. \cite{miller}. As the computation is performed in a conservative system, it will have its own advantages like re-usage of the output solitons.  Another important point that can be inferred from the above construction is that the collision in the Manakov system can be dynamically reconfigured to realize NOR gate.  This successful theoretical construction of two input optical logic gate by exploiting the energy sharing collisions of the Manakov solitons suggests the possibility of employing the very same idea to implement quantum logic gates such as X-gate, Y-gate, Z-gate and Hadamard gate. Our work can be extended to realize multi-state logic too by considering multicomponent nonlinear Schr\"odinger system with more than two components. Work is in progress in these directions.
\begin{acknowledgments}
M. V. acknowledges the support of Science and Engineering Research Board, Department of Science and Technology (DST-SERB), Government of India, Start Up Research Grant (Young Scientist: File No.YSS/2015/000629). The work of T. K. is supported by Science and Engineering Research Board, Department of Science and Technology (DST-SERB), Government of India, in the form of a major research project (File No. EMR/2015/001408). The work of M. L. is supported by a DST-SERB Distinguished Fellowship Programme and a DST/SERB project (Diary No.SERB/F/4307/2016-17). The authors thank the referees for their critical comments and for providing a few important references which helped them to present the material in a proper perspective.
\end{acknowledgments}	

\end{document}

%\preprint{}

\title{Harnessing energy sharing collisions of Manakov solitons to implement universal NOR and OR logic gates - Supplemental material}


\author{M. Vijayajayanthi\footnote{ e-mail: vijayajayanthi.cnld@gmail.com}}
%\thanks{}
\affiliation{Department of Physics,
 \\B. S. Abdur Rahman Crescent Institute of Science and Technology,
 \\ Chennai--600 048, India}
\author{T. Kanna\footnote{ e-mail: kanna$\_$phy@bhc.edu.in}}
\affiliation{PG and Research Department of Physics,
 \\
Bishop Heber College,Tiruchirapalli--620 017, India}
\author{K. Murali\footnote{ e-mail: kmurali@annauniv.edu}}
\affiliation{Department of Physics, Anna University, Chennai--600 025, India}
\author{M. Lakshmanan\footnote{ e-mail: lakshman@cnld.bdu.ac.in}}
\affiliation{Centre for Nonlinear Dynamics, School of Physics, Bharathidasan University, Tiruchirapalli--620 024, India\\}

\date{\today}

%\begin{abstract}
%The fascinating energy sharing collision of bright optical solitons in the celebrated Manakov system, governing pico-second pulse propagation in randomly varying high birefringent fiber is profitably employed to realize two input logic gates.  Especially, here the universal NOR gate and the OR gate are constructed from the energy sharing collision of four bright solitons which is well described by the exact bright four-soliton solution of the Manakov system.  The experimental observation of such energy sharing collisions of the Manakov solitons ensures the possibility of experimental constrction of such gates.
%\end{abstract}

\pacs{}
%\keywords{}

\maketitle


\section*{Four soliton solution of the Manakov system}
Using the Hirota's bilinearization method, we obtain the four bright soliton solution of the Manakov system in Gram determinant form as below \cite{kannapre,epj}:
%\begin{subequations}
\bea
q_s=\frac{g^{(s)}}{f}, \quad s=1,2.\nonumber
\eea
where
\bea
g^{(s)}=
\left|
\begin{array}{ccc}
A & I & \phi\\
-I & B & {\bf 0}^T\\
{\bf 0} & C_s & 0
\end{array}
\right|,
\quad
f= \left|
\begin{array}{cc}
A & I\\
-I & B
\end{array}
\right|.\nonumber
\eea
In the above expression,  $I$ is a $(4\times 4)$ identity matrix,
%\begin{subequations}
\bea
C_s= -\left(\alpha_1^{(s)}, \alpha_2^{(s)},  \alpha_{3}^{(s)},  \alpha_{4}^{(s)}\right),\quad  {\bf{0}}=(0, 0,  0, 0).\;\;\;
%\eea
%\bea
\psi_j=\left(
\begin{array}{c}
\alpha_j^{(1)}\\
\alpha_j^{(2)}\\
\end{array}
\right), \quad \phi=\left(
\begin{array}{c}
e^{\eta_1}\\
e^{\eta_2}\\
e^{\eta_3}\\
e^{\eta_{4}}
\end{array}
\right),\nonumber
\eea
The elements of the matrices $A$ and $B$ are given by
\bea
A_{ij}= \frac{e^{\eta_i+\eta_j^*}}{(k_i+k_j^*)},  \quad B_{ij}=\kappa_{ji}=\frac{\psi_i^{\dagger}  \psi_j}{(k_i^*+k_j)}, \quad i,j=1, 2, \ldots, 4.  \label{omg}\nonumber
\eea
%\end{subequations}
%\noindent where $A_{ij}=\ds{\frac{e^{\eta_i+\eta_j^*}}{k_i+k_j^*}}$, and $B_{ij}=\kappa_{ji}=\ds{\frac{\left(\sum_{s=1}^{N}\alpha_j^{(s)}\alpha_i^{(s)*}\right)}{(k_j+k_i^*)}}, i,j=1,2,3,4$.
In equation (\ref{omg}),  $\dagger$ represents the transpose conjugate, $k_j=k_{jR}+i k_{jI}, j=1,2,$, where the real part of $k_j$ ($k_{jR}$) represent the amplitudes of the solitons and the complex part of $k_j$ ($k_{jI}$) represent the velocities of the solitons.  One can refer  to \cite{epj} for a detailed derivation of the above Gram determinant form of four soliton solution.
\newpage
\section*{Schematic of four soliton collision process}
The schematic pair-wise four soliton collision process considered in our work is shown below.
\begin{figure}[ht]
\includegraphics[width=0.3\linewidth]{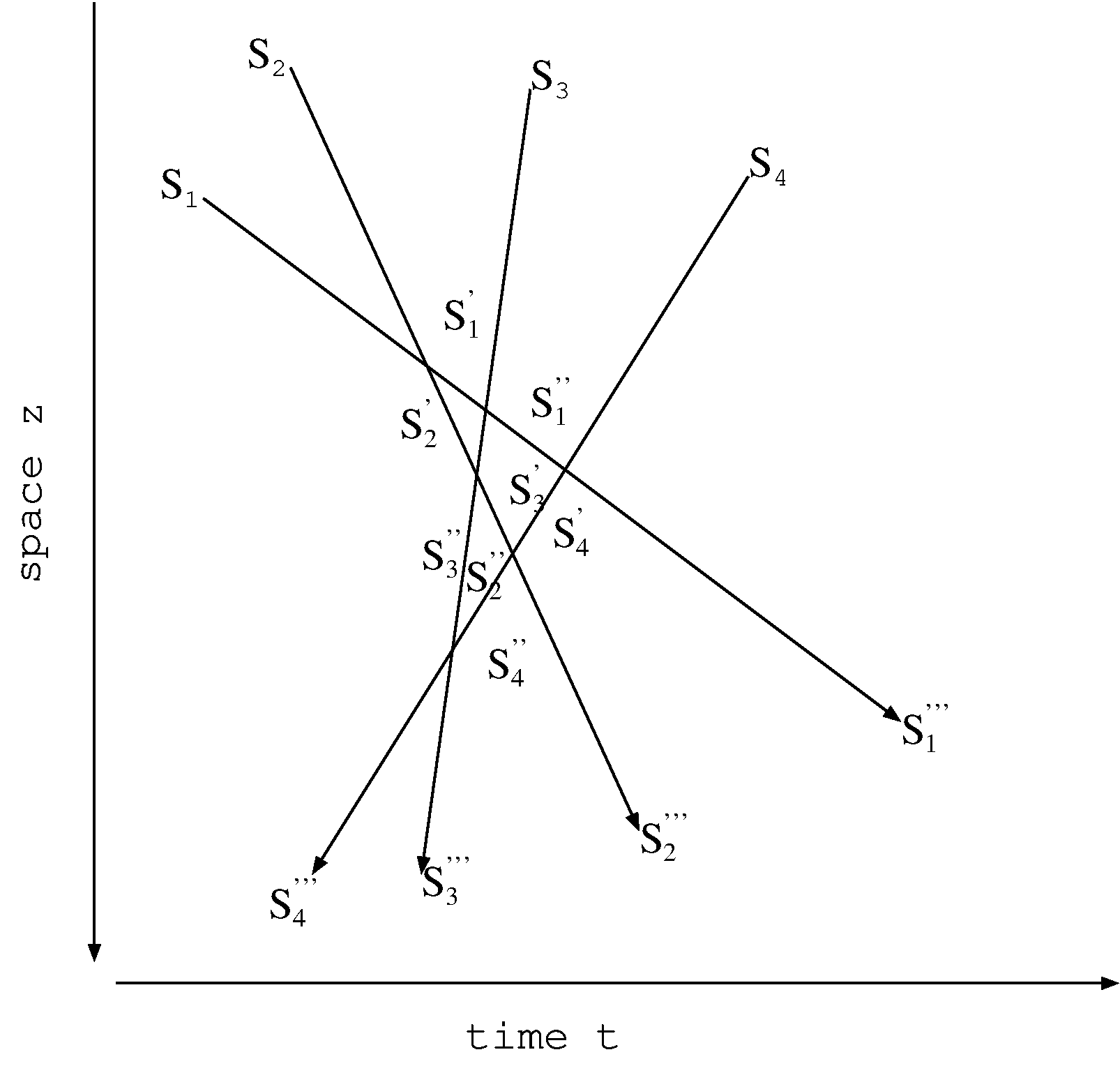}
\caption{Collision picture of solitons $S_1$, $S_2$, $S_3$ and $S_4$.}\label{collision}
\end{figure}
\section*{Asymptotic analysis of four soliton solution of the Manakov system}
Considering the above four soliton solution,
without loss of generality, we assume that the
quantities $k_{1R}$,
$k_{2R}$,$k_{3R}$,  and $k_{4R}$ are positive and  $k_{1I}>k_{2I}>k_{3I}>k_{4I}$ .
For this condition, the asymptotic behaviour variables $\eta_{iR}$'s, $i=1,2,3,4,$
for the four solitons ($S_1$, $S_2$, $S_3$ and $S_4$) is given below.

(i) $\eta_{1R} \approx 0$,  $\eta_{2R} \rightarrow \pm \infty$,
$\eta_{3R} \rightarrow \pm \infty$, $\eta_{4R} \rightarrow \pm \infty$, as $z \rightarrow \pm \infty$,

(ii) $\eta_{2R} \approx 0$,  $\eta_{1R} \rightarrow \mp \infty$,
$\eta_{3R} \rightarrow \pm \infty$, $\eta_{4R} \rightarrow \pm \infty$, as $z \rightarrow \pm \infty$,

(iii) $\eta_{3R} \approx 0$,  $\eta_{1R} \rightarrow \mp \infty$,
$\eta_{2R} \rightarrow \mp \infty$, $\eta_{4R} \rightarrow \pm \infty$, as $z \rightarrow \pm \infty$,

(iv) $\eta_{4R} \approx 0$,  $\eta_{1R} \rightarrow \mp \infty$,
$\eta_{2R} \rightarrow \mp \infty$,  $\eta_{3R} \rightarrow \mp \infty$, as $z \rightarrow \pm \infty$.

We have the following asymptotic forms of the above four-soliton solution.\\
\noindent\underline{(i)Before Collision (limit $ z \rightarrow -\infty$)}:\\
\noindent(a) \underline{\it Soliton 1}  ($\eta_{1R} \approx 0$, $\eta_{2R} \rightarrow -\infty$,
$\eta_{3R} \rightarrow -\infty$, $\eta_{4R} \rightarrow -\infty$):
%\begin{subequations}
\begin{eqnarray}
\left(
\begin{array}{c}
q_1\\
q_2
\end{array}
\right)
&\approx &
\left(
\begin{array}{c}
A_1^{1-} \\
A_2^{1-}
\end{array}
\right)k_{1R}
{\mbox{sech}\,\left(\eta_{1R}+\frac{R_1}{2}\right)}e^{i\eta_{1I}} ,\nonumber\\
\left(
\begin{array}{c}
A_1^{1-} \\
A_2^{1-}
\end{array}
\right)
&=&
\left(
\begin{array}{c}
\alpha_1^{(1)}\\
\alpha_1^{(2)}
\end{array}
\right) \frac{e^{\frac{-R_1}{2}}}{(k_1+k_1^*)}.\nonumber
\eea
%\ees
\noindent(b) \underline{\it Soliton 2}  ($\eta_{2R} \approx 0$, $\eta_{1R} \rightarrow \infty$,
$\eta_{3R} \rightarrow -\infty$, $\eta_{4R} \rightarrow -\infty$):
%\bes
\begin{eqnarray}
\left(
\begin{array}{c}
q_1\\
q_2
\end{array}
\right)
&\approx&
\left(
\begin{array}{c}
A_1^{2-} \\
A_2^{2-}
\end{array}
\right)k_{2R}
{\mbox{sech}\,\left(\eta_{2R}+\frac{R_4-R_1}{2}\right)}e^{i\eta_{2I}} ,\nonumber\\
\left(
\begin{array}{c}
A_1^{2-}\\
A_2^{2-}
\end{array}
\right)
&=&
\left(
\begin{array}{c}
e^{\delta_{11}}\\
e^{\delta_{12}}
\end{array}
\right)\frac{e^{-\frac{(R_1+R_4)}{2}}}{(k_2+k_2^*)}.\nonumber
\eea
%\ees

\noindent (c) \underline{\it Soliton 3}  ($\eta_{3R} \approx 0$, $\eta_{1R} \rightarrow \infty$,
$\eta_{2R} \rightarrow \infty$, $\eta_{4R} \rightarrow -\infty$):
%\bes
\begin{eqnarray}
\left(
\begin{array}{c}
q_1\\
q_2
\end{array}
\right)
&\approx&
\left(
\begin{array}{c}
A_1^{3-} \\
A_2^{3-}
\end{array}
\right)k_{3R}
{\mbox{sech}\,\left(\eta_{3R}+\frac{R_7-R_4}{2}\right)}e^{i\eta_{3I}} ,\nonumber\\
\left(
\begin{array}{c}
A_1^{3-}\\
A_2^{3-}
\end{array}
\right)
&=&
\left(
\begin{array}{c}
e^{\tau_{11}}\\
e^{\tau_{12}}
\end{array}
\right)\frac{e^{-\frac{(R_4+R_7)}{2}}}{(k_3+k_3^*)}.\nonumber
\eea
%\end{subequations}

\noindent(d) \underline{\it Soliton 4}  ($\eta_{4R} \approx 0$, $\eta_{1R} \rightarrow \infty$,
$\eta_{2R} \rightarrow \infty$, $\eta_{3R} \rightarrow \infty$):
%\begin{subequations}
\begin{eqnarray}
\left(
\begin{array}{c}
q_1\\
q_2
\end{array}
\right)
&\approx &
\left(
\begin{array}{c}
A_1^{4-} \\
A_2^{4-}
\end{array}
\right)k_{4R}
{\mbox{sech}\,\left(\eta_{4R}+\phi^{4-}\right)}e^{i\eta_{4I}},\nonumber\\
\left(
\begin{array}{c}
A_1^{4-} \\
A_2^{4-}
\end{array}
\right)
&=&
-\sqrt{\frac{n^{4-}}{2 k_{4R}(n^{4-})^{*}}}\left(
\begin{array}{c}
N_1^{4-}\\
N_2^{4-}
\end{array}
\right) \frac{1}{(D_1^{4-}D_2^{4-})^{1/2}},\nonumber
\eea
In the above equations, the various other quantities are defined below:
%\begin{subequations}
\bea
e^{\delta_{1j}}=\frac{(k_1-k_2)(\alpha_1^{(j)}\kappa_{21}-\alpha_2^{(j)}\kappa_{11}
)}{(k_1+k_1^*)(k_1^*+k_2)},\;\;j=1,2,\nonumber
\eea
%e^{\delta_{6j}}=\frac{(k_2-k_3)(\alpha_2^{(j)}\kappa_{33}-\alpha_3^{(j)}\kappa_{23}
%)}{(k_3^*+k_2)(k_3^*+k_3)},\nonumber\;\;j=1,2,\\
\bea
&&e^{\tau_{1j}}=\frac{(k_2-k_1)(k_3-k_1)(k_3-k_2)(k_2^*-k_1^*)}
{(k_1^*+k_1)(k_1^*+k_2)(k_1^*+k_3)(k_2^*+k_1)(k_2^*+k_2)(k_2^*+k_3)}\nonumber\\
&&\times
\left[\alpha_1^{(j)}(\kappa_{21}\kappa_{32}-\kappa_{22}\kappa_{31})
+\alpha_2^{(j)}(\kappa_{12}\kappa_{31}-\kappa_{32}\kappa_{11})
+\alpha_3^{(j)}(\kappa_{11}\kappa_{22}-\kappa_{12}\kappa_{21})
\right],\nonumber
\eea
%\vspace{-0.7cm}
\bea
e^{R_1}=\frac{\kappa_{11}}{k_1+k_1^*}, \quad
e^{R_4}=\frac{(k_2-k_1)(k_2^*-k_1^*)}
{(k_1^*+k_1)(k_1^*+k_2)(k_1+k_2^*)(k_2^*+k_2)}
\left[\kappa_{11}\kappa_{22}-\kappa_{12}\kappa_{21}\right],\nonumber
\eea
\bea
&&e^{R_7}= \frac{|k_1-k_2|^2|k_2-k_3|^2|k_3-k_1|^2}
{(k_1+k_1^*)(k_2+k_2^*)(k_3+k_3^*)|k_1+k_2^*|^2|k_2+k_3^*|^2|k_3+k_1^*|^2}
\nonumber\\
&&\times\left[(\kappa_{11}\kappa_{22}\kappa_{33}-
\kappa_{11}\kappa_{23}\kappa_{32})
+(\kappa_{12}\kappa_{23}\kappa_{31}-
\kappa_{12}\kappa_{21}\kappa_{33})+(\kappa_{21}\kappa_{13}\kappa_{32}-
\kappa_{22}\kappa_{13}\kappa_{31})\right],\nonumber
\eea
\bea
&&n^{4-}=(k_4-k_1)(k_4-k_2)(k_4-k_3)(k_1+k_4^*)(k_2+k_4^*)(k_3+k_4^*),\nonumber\\
&&N_j^{4-}=\left|
\begin{array}{cccc}
\alpha_1^{(j)}&\alpha_2^{(j)}&\alpha_3^{(j)}&\alpha_4^{(j)}\\
\kappa_{11} &\kappa_{21} &\kappa_{31}&\kappa_{41}\\
\kappa_{12} &\kappa_{22}&\kappa_{32}&\kappa_{42}\\
\kappa_{13} &\kappa_{23}&\kappa_{33}&\kappa_{43}\\
\end{array}
\right|,\quad j=1, 2,\nonumber\\
&&D_1^{4-}=\left|
\begin{array}{ccc}
\kappa_{11} &\kappa_{12} &\kappa_{13}\\
\kappa_{21} &\kappa_{22}&\kappa_{23}\\
\kappa_{31} &\kappa_{32}&\kappa_{33}\\
\end{array}
\right|,\quad D_2^{4-}=\left|
\begin{array}{cccc}
\kappa_{11} &\kappa_{12} &\kappa_{13}&\kappa_{14}\\
\kappa_{21} &\kappa_{22}&\kappa_{23}&\kappa_{24}\\
\kappa_{31} &\kappa_{32}&\kappa_{33}&\kappa_{34}\\
\kappa_{41} &\kappa_{42}&\kappa_{43}&\kappa_{44}\\
\end{array}
\right|,\nonumber\\
&&\phi^{4-}
=
\ln\Bigg[\frac{|k_4-k_1||k_2-k_4||k_3-k_4|}{|k_1+k_4^*||k_2+k_4^*||k_3+k_4^*|\sqrt{2 k_{4R}}}\Bigg]+\frac{1}{2}\ln\Bigg[\frac{D_2^{4-}}{D_1^{4-}}\Bigg],\nonumber
\eea
and
\bea
\kappa_{il}= \frac{\sum_{n=1}^2\alpha_i^{(n)}\alpha_l^{(n)*}}
{\left(k_i+k_l^*\right)},\;i,l=1,2,3,4.\nonumber
\eea
%\end{subequations}
%\ees
\noindent\underline{(ii)After Collision (limit $ z \rightarrow \infty$)}:\\
After interaction (as $ z \rightarrow \infty$) the forms of the solitons are given below.\\
\noindent(a) \underline{\it Soliton 1}  ($\eta_{1R} \approx 0$, $\eta_{2R} \rightarrow \infty$,
$\eta_{3R} \rightarrow \infty$, $\eta_{4R} \rightarrow \infty$):
%\begin{subequations}
\begin{eqnarray}
\left(
\begin{array}{c}
q_1\\
q_2
\end{array}
\right)
&\approx &
\left(
\begin{array}{c}
A_1^{1+} \\
A_2^{1+}
\end{array}
\right)k_{1R}
{\mbox{sech}\,\left(\eta_{1R}+\phi^{1+}\right)}e^{i\eta_{1I}},\nonumber
\eea
where
\bea
\left(
\begin{array}{c}
A_1^{1+} \\
A_2^{1+}
\end{array}
\right)
&=&
-\sqrt{\frac{n^{1+}}{2 k_{1R}(n^{1+})^{*}}}\left(
\begin{array}{c}
N_1^{1+}\\
N_2^{1+}
\end{array}
\right) \frac{1}{(D_1^{1+}D_2^{4-})^{1/2}},\nonumber\\
\phi^{1+}
&=&
\ln\Bigg[\frac{|k_1-k_2||k_4-k_1||k_3-k_1|}{|k_1+k_2^*||k_1+k_3^*||k_1+k_4^*|\sqrt{2 k_{1R}}}\Bigg]+\frac{1}{2}\ln\Bigg[\frac{D_2^{4-}}{D_1^{1+}}\Bigg].\nonumber
\eea
In the above equations,
\bea
n^{1+}&=&(k_4-k_1)(k_3-k_1)(k_2-k_1)(k_2+k_1^*)(k_3+k_1^*)(k_4+k_1^*),\nonumber\\
N_j^{1+}&=&\left|
\begin{array}{cccc}
\alpha_1^{(j)}&\alpha_2^{(j)}&\alpha_3^{(j)}&\alpha_4^{(j)}\\
\kappa_{12} &\kappa_{22}&\kappa_{32}&\kappa_{42}\\
\kappa_{13} &\kappa_{23}&\kappa_{33}&\kappa_{43}\\
\kappa_{14} &\kappa_{24} &\kappa_{34}&\kappa_{44}\\
\end{array}
\right|, j=1, 2, \quad
D_1^{1+}=\left|
\begin{array}{ccc}
\kappa_{22} &\kappa_{23} &\kappa_{24}\\
\kappa_{32} &\kappa_{33}&\kappa_{34}\\
\kappa_{42} &\kappa_{43}&\kappa_{44}\\
\end{array}
\right|.\nonumber
\eea
%\end{subequations}
\noindent(b) \underline{\it Soliton 2}  ($\eta_{2R} \approx 0$, $\eta_{1R} \rightarrow -\infty$,
$\eta_{3R} \rightarrow \infty$, $\eta_{4R} \rightarrow \infty$):
%\begin{subequations}
\begin{eqnarray}
\left(
\begin{array}{c}
q_1\\
q_2
\end{array}
\right)
&\approx &
\left(
\begin{array}{c}
A_1^{2+} \\
A_2^{2+}
\end{array}
\right)k_{2R}
{\mbox{sech}\,\left(\eta_{2R}+\phi^{2+}\right)}e^{i\eta_{2I}},\nonumber
\eea
where
\bea
\left(
\begin{array}{c}
A_1^{2+} \\
A_2^{2+}
\end{array}
\right)
&=&
\frac{\sqrt{\kappa_{22}}}{\sqrt{\mu \bigg(|\alpha_2^{(1)}|^2+|\alpha_2^{(2)}|^2}\bigg)}\left(
\begin{array}{c}
\frac{N_1^{1+}}{\alpha_2^{(1)}}\\
\frac{N_2^{1+}}{\alpha_2^{(2)}}
\end{array}
\right)\left(
\frac{
f_1}{
f_1^*}
\right)\left(
\begin{array}{c}
\alpha_2^{(1)}\\
\alpha_2^{(2)}
\end{array}
\right)
\frac{1}{(D_2^{1+}D_2^{2+})^{1/2}},\nonumber\\
\phi^{2+}
&=&
\ln\Bigg[\frac{|k_2-k_3||k_2-k_4|}{|k_2+k_3^*||k_4+k_2^*|\sqrt{2 k_{2R}}}\Bigg]+\frac{1}{2}\ln\Bigg[\frac{D_2^{2+}}{D_2^{1+}}\Bigg].\nonumber
\eea
%\end{subequations}
The quantities $f_1$, $N_1^{2+},$ $N_2^{2+},$ $D_2^{1+}$, and $D_2^{2+}$ are given below:
\bea
f_1&=&\sqrt{(k_4+k_2^*)(k_2^*+k_3)(k_3-k_2)(k_4-k_2)},\nonumber\\
N_1^{2+}&=&\left|
\begin{array}{ccc}
\alpha_2^{(1)} &\alpha_3^{(1)} &\alpha_4^{(1)}\\
\kappa_{23} &\kappa_{33}&\kappa_{43}\\
\kappa_{24} &\kappa_{34}&\kappa_{44}\\
\end{array}
\right|,\quad N_2^{2+}=\left|
\begin{array}{ccc}
\alpha_2^{(2)} &\alpha_3^{(2)} &\alpha_4^{(2)}\\
\kappa_{23} &\kappa_{33}&\kappa_{43}\\
\kappa_{24} &\kappa_{34}&\kappa_{44}\\
\end{array}
\right|,\nonumber\\
D_2^{1+}&=&\left|
\begin{array}{cc}
\kappa_{33} &\kappa_{34} \\
\kappa_{43} &\kappa_{44}\\
\end{array}
\right|,\quad D_2^{2+}=\left|
\begin{array}{ccc}
\kappa_{22} &\kappa_{32} &\kappa_{42}\\
\kappa_{23} &\kappa_{33}&\kappa_{43}\\
\kappa_{24} &\kappa_{34}&\kappa_{44}\\
\end{array}
\right|.\nonumber
\eea
\noindent(c) \underline{\it Soliton 3}  ($\eta_{3R} \approx 0$, $\eta_{1R} \rightarrow -\infty$,
$\eta_{2R} \rightarrow -\infty$, $\eta_{4R} \rightarrow \infty$):
%\begin{subequations}
\begin{eqnarray}
\left(
\begin{array}{c}
q_1\\
q_2
\end{array}
\right)
&\approx &
\left(
\begin{array}{c}
A_1^{3+} \\
A_2^{3+}
\end{array}
\right)k_{3R}
{\mbox{sech}\,\left(\eta_{3R}+\phi^{3+}\right)}e^{i\eta_{3I}},\nonumber
\eea
where
\bea
\left(
\begin{array}{c}
A_1^{3+} \\
A_2^{3+}
\end{array}
\right)
&=&
\frac{1}{\sqrt{\mu \bigg(|\alpha_3^{(1)}|^2+|\alpha_3^{(2)}|^2}\bigg)}\left(
\begin{array}{c}
\frac{N_1^{3+}}{D_1^{3+}}\\
\frac{N_2^{3+}}{D_1^{3+}}
\end{array}
\right)\left(
\frac
{g_2}
{g_2^*}
\right)\Bigg(\frac{\kappa_{43}\kappa_{33}}{\kappa_{34}\kappa_{44}}\Bigg)^{\frac{1}{2}}
,\nonumber\\
\phi^{3+}
&=&
\ln\Bigg[\frac{|k_4-k_3|}{|k_3+k_4^*|\sqrt{2 k_{3R}}}\Bigg]+\ln\Bigg[\frac{D_1^{3+}}{\sqrt{\kappa_{44}}}\Bigg],\nonumber
\eea
%\end{subequations}
Here
\bea
g_2&=&(k_3^*+k_4)\sqrt{(k_3-k_4) (\alpha_3^{(1)}\alpha_4^{(1)*}+\alpha_3^{(2)}\alpha_4^{(2)*})},\nonumber\\
N_1^{3+}&=&\left|
\begin{array}{cc}
\alpha_3^{(1)} &\alpha_4^{(1)} \\
\kappa_{34} &\kappa_{44}\\
\end{array}
\right|,\quad N_2^{3+}=\left|
\begin{array}{cc}
\alpha_3^{(2)} &\alpha
_4^{(2)} \\
\kappa_{34} &\kappa_{44}\\
\end{array}\right|,\nonumber\\
D_1^{3+}&=&\left|
\begin{array}{cc}
\kappa_{33} &\kappa_{34} \\
\kappa_{43} &\kappa_{44}\\
\end{array}
\right|^\frac{1}{2}.\nonumber
\eea
\noindent(d) \underline{\it Soliton 4}  ($\eta_{4R} \approx 0$, $\eta_{1R} \rightarrow -\infty$,
$\eta_{2R} \rightarrow -\infty$, $\eta_{3R} \rightarrow -\infty$):
%\begin{subequations}
\begin{eqnarray}
\left(
\begin{array}{c}
q_1\\
q_2
\end{array}
\right)
&\approx &
\left(
\begin{array}{c}
A_1^{4+} \\
A_2^{4+}
\end{array}
\right)k_{4R}
{\mbox{sech}\,\left(\eta_{4R}+\phi^{4+}\right)}e^{i\eta_{4I}}\nonumber,
\eea
where
\bea
\left(
\begin{array}{c}
A_1^{4+} \\
A_2^{4+}
\end{array}
\right)
&=&
\left(
\begin{array}{c}
\alpha_4^{(1)}\\
\alpha_4^{(2)}
\end{array}
\right)\frac{e^{D_1^{4+}/2}}{(k_4+k_4^*)},\nonumber\\
\phi^{4+}
&=&
\frac{D_1^{4+}}{2}, \quad e^{D_1^{4+}}=\frac{\mu (|\alpha_4^{(1)}|^2+|\alpha_4^{(2)}|^2)} {(k_4+k_4^*)^2}.\nonumber
\eea
%\end{subequations}
\newpage
\section{Two dimensional plots of intensities of NOR gate}
The Figures (2-4) of columns (a) and (b) depict the two dimensional plots of intensities of the NOR gate at the input $(z=-15)$ and at the output $(z=15)$, respectively corresponding to the figures  2(a) - 2(c) in the main text.
\begin{figure}[h]
\begin{center}
\includegraphics[width=0.6\linewidth]{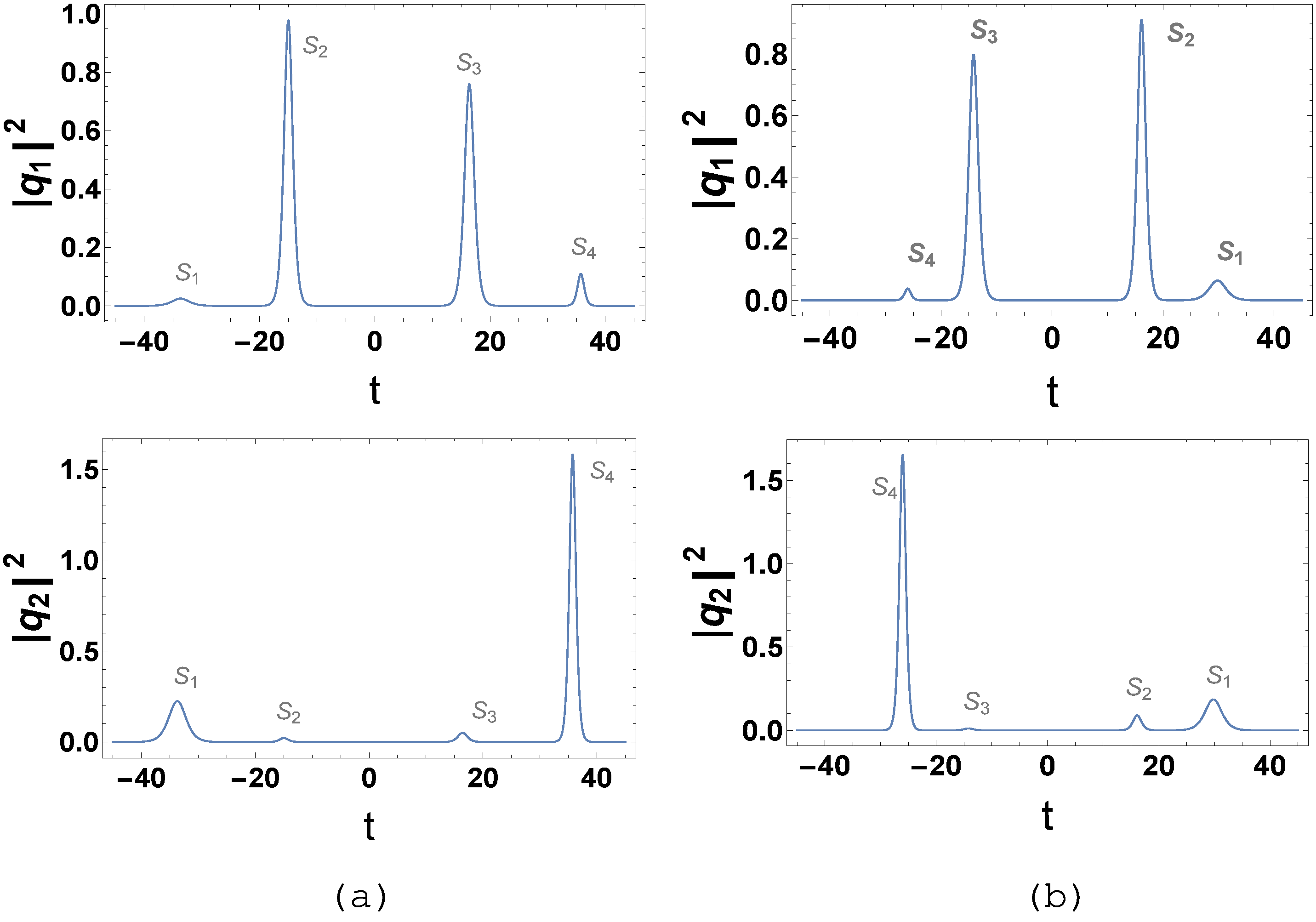}
\caption{NOR gate: Input states  $``0"$ and  $``1"$; Output state  $``0"$.}\label{nor_01_2d}
\end{center}
\end{figure}
\begin{figure}[h]
\begin{center}
\includegraphics[width=0.6\linewidth]{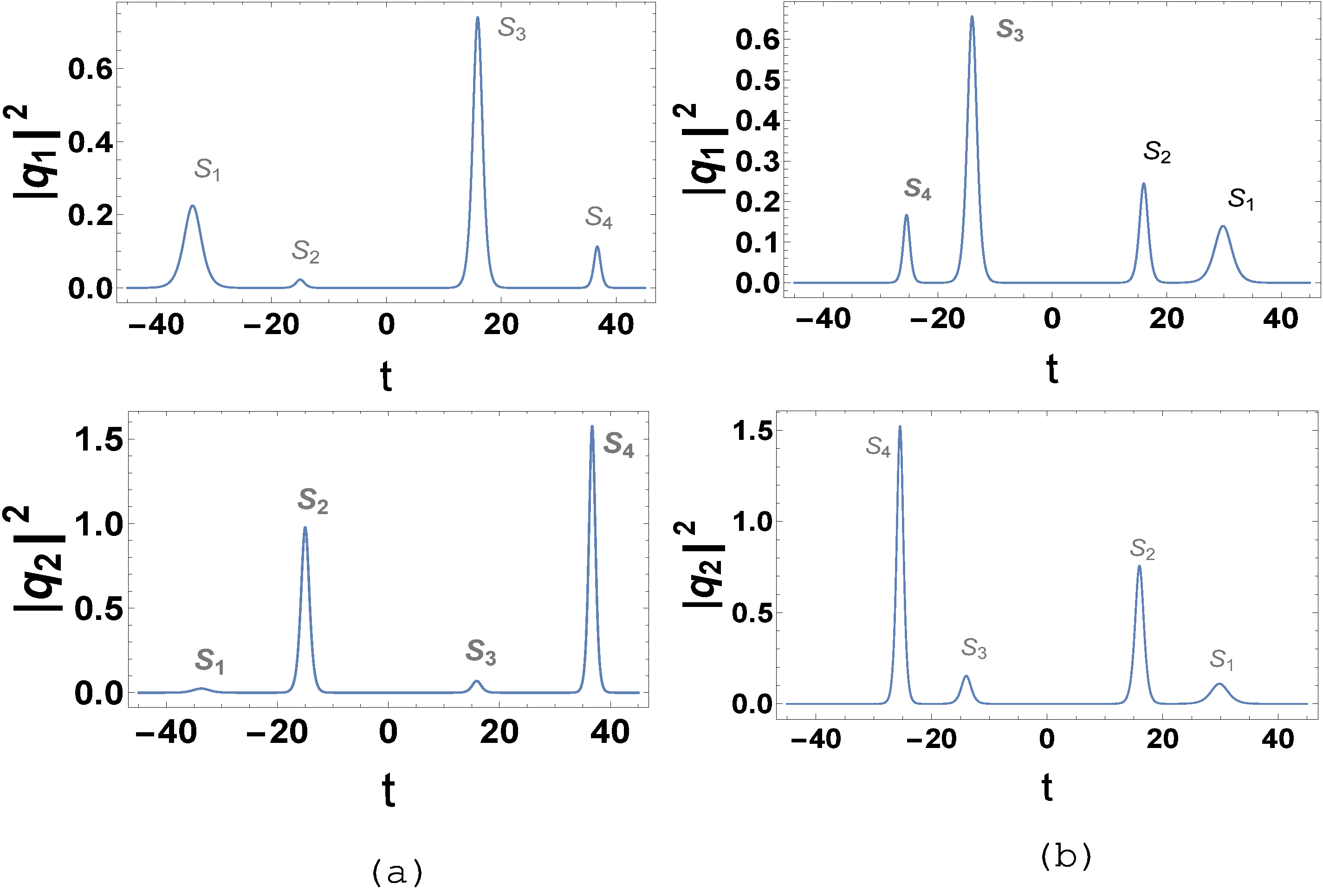}
\caption{NOR gate:  Input states  $``1"$ and  $``0"$; Output state  $``0"$. }\label{nor_10_2d}
\end{center}
\end{figure}
\begin{figure}[h]
\begin{center}
\includegraphics[width=0.6\linewidth]{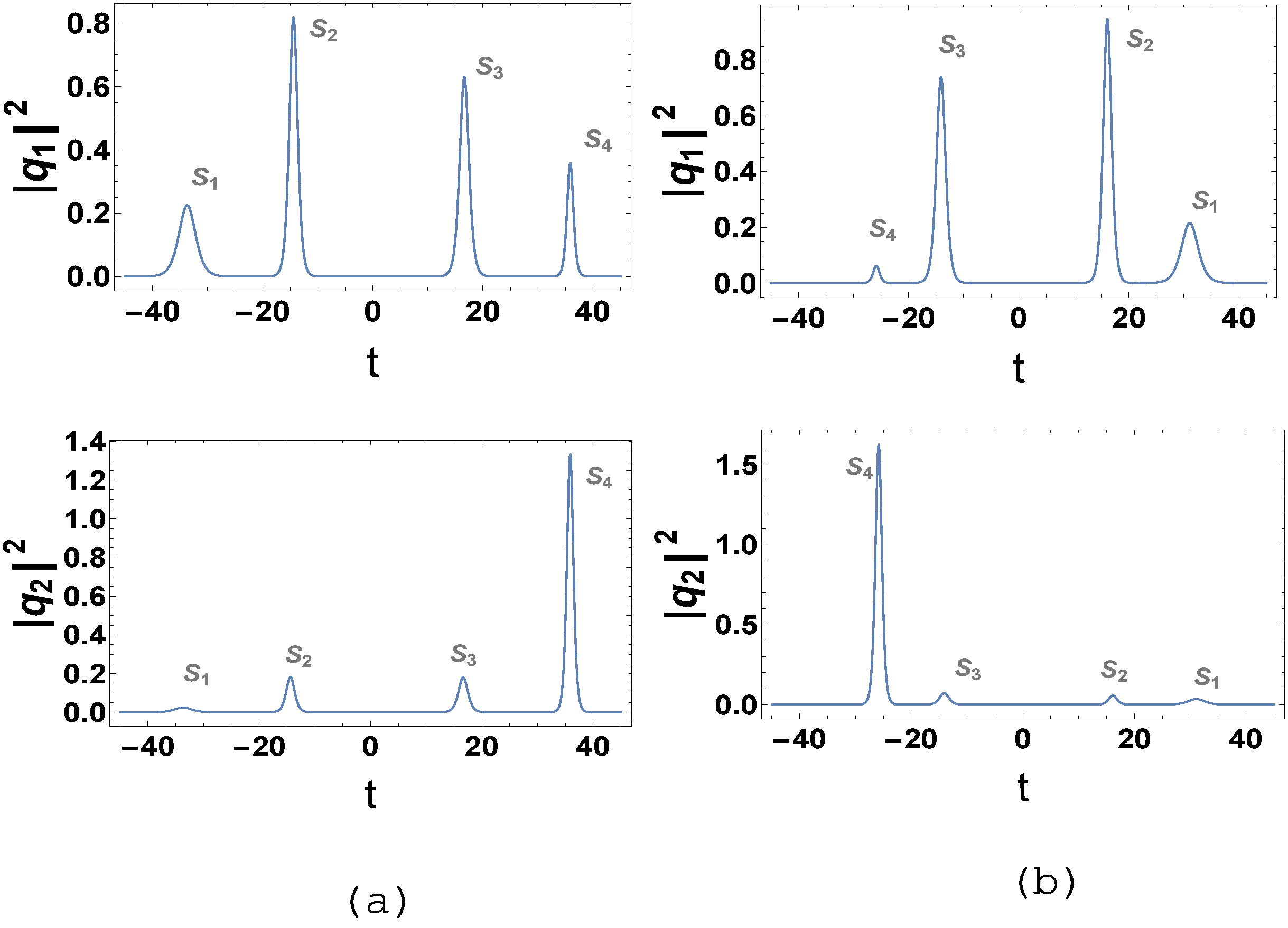}
\caption{NOR gate: Input states  $``1"$ and  $``1"$; Output state  $``0"$. }\label{nor_11_2d}
\end{center}
\end{figure}

\section{Explicit construction of OR gate}
To construct the OR gate,  two inputs are fed into the solitons $S_1$ and $S_2$ and the output is taken up from soliton $S_4$ . In order to get the desired output satisfying the truth table of the OR gate, we make use of the condition $\rho^{4+}=\rho^{1-}+\rho^{2-}$ and choose
\bea
\alpha_4^{(1)}= \left(\frac{\alpha_1^{(1)}}{\alpha_1^{(2)}}+\frac{\alpha_2^{(1)}\left((k_1-k_2)|\alpha_1^{(1)}|^2-(k_2+k_1^*)|
\alpha_1^{(2)}|^2\right)+(k_1+k_1^*)\alpha_1^{(1)}\alpha_1^{(2)*}\alpha_2^{(2)}}{(k_1+k_1^*)\alpha_1^{(1)*}\alpha_2^{(1)}\alpha_1^{(2)}-\alpha_2^{(2)}\left((k_2-k_1)
|\alpha_1^{(2)}|^2+(k_2+k_1^*)|\alpha_1^{(1)}|^2\right)}\right) \alpha_4^{(2)}\nonumber.
\eea
% where $R$ is defined below Eq. (5) of main paper.
 To achieve the input states of OR gate such as $``0" \&``0"$,  $``0" \& ``1"$,  $``1" \& ``0"$, and  $``1" \& ``1"$, the polarization parameters $(\alpha_{l}^{(m)},l=m=1,2)$ of solitons $S_1$ and $S_2$ are chosen as the same as that of the NOR gate.  Similarly, all the other soliton parameters are chosen as in the case of the NOR gate except with $\alpha_4^{(2)}=0.001-0.002 i$.   Figures (5-8) depict the implementation of the OR gate.  One can refer tables I and II as the truth table and the intensity table of the OR gate, respectively.
\begin{figure}
\begin{center}
\includegraphics[width=1\linewidth]{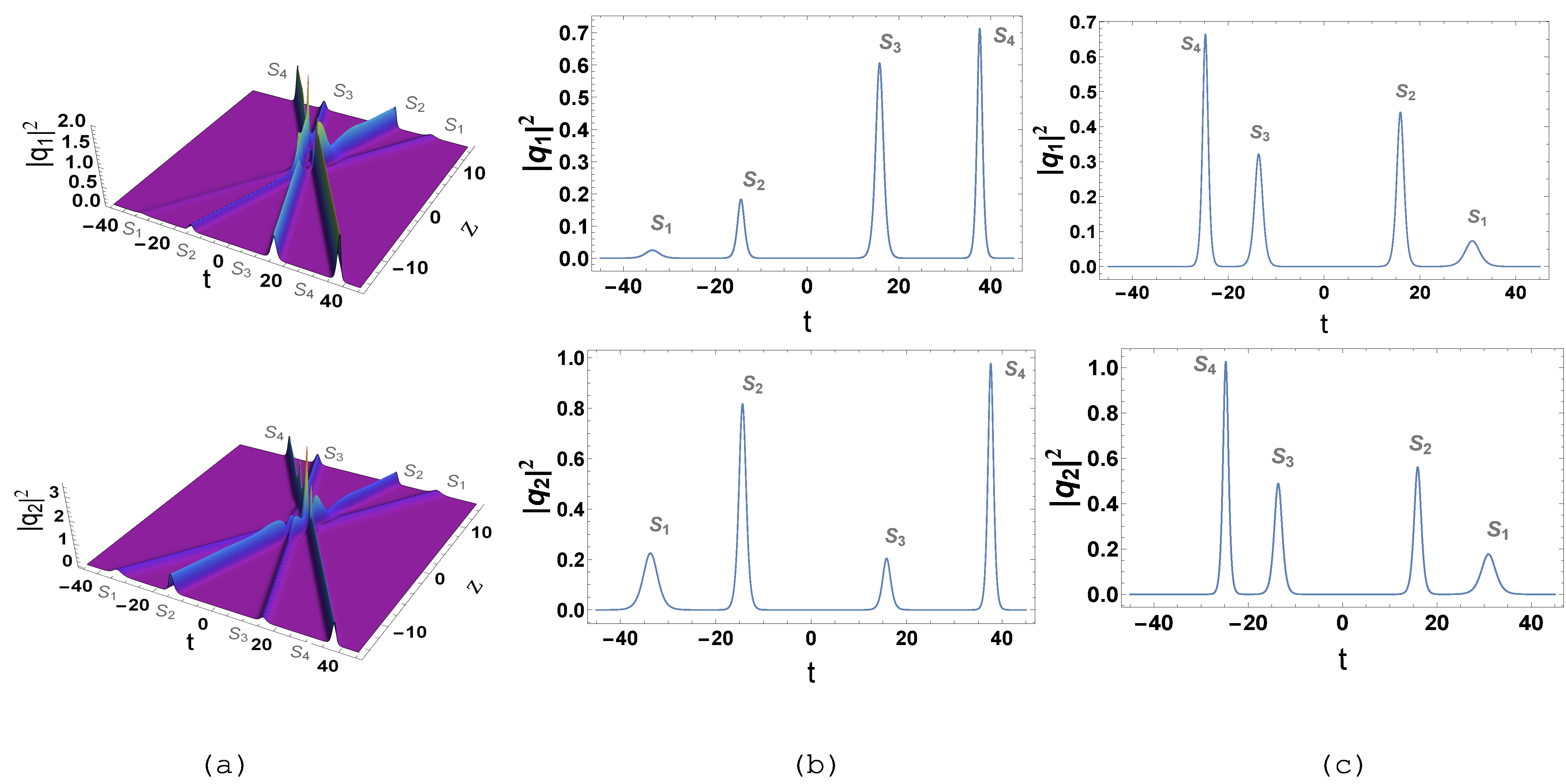}
\caption{OR gate: Input states  $``0"$ and  $``0"$; Output state  $``0"$. }\label{or_00}
\end{center}
%\end{figure}
%\begin{figure}
\begin{center}
\includegraphics[width=1 \linewidth]{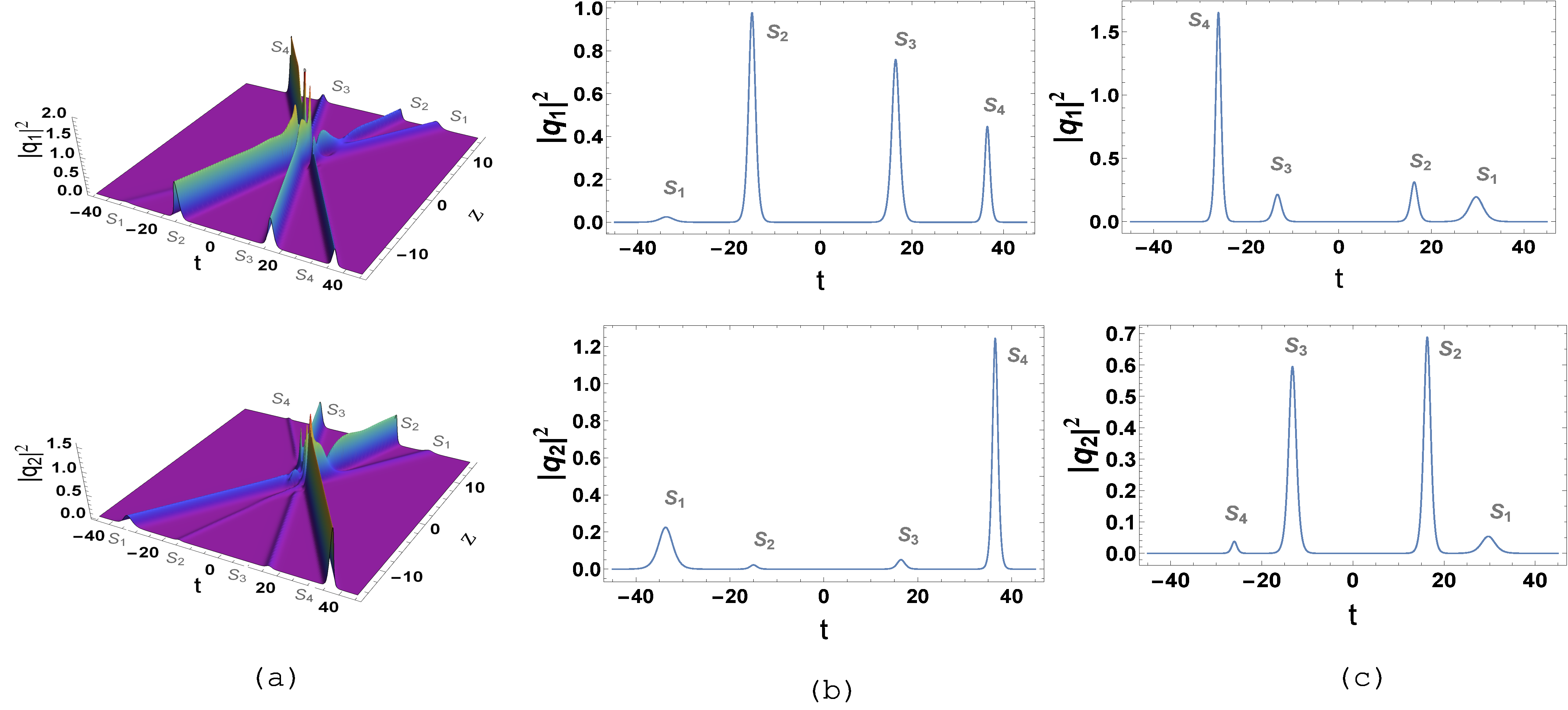}
\caption{OR gate: Input states  $``0"$ and  $``1"$; Output state  $``1"$. }\label{or_01}
\end{center}
\end{figure}
\begin{figure}[!ht]
\begin{center}
\includegraphics[width=1 \linewidth]{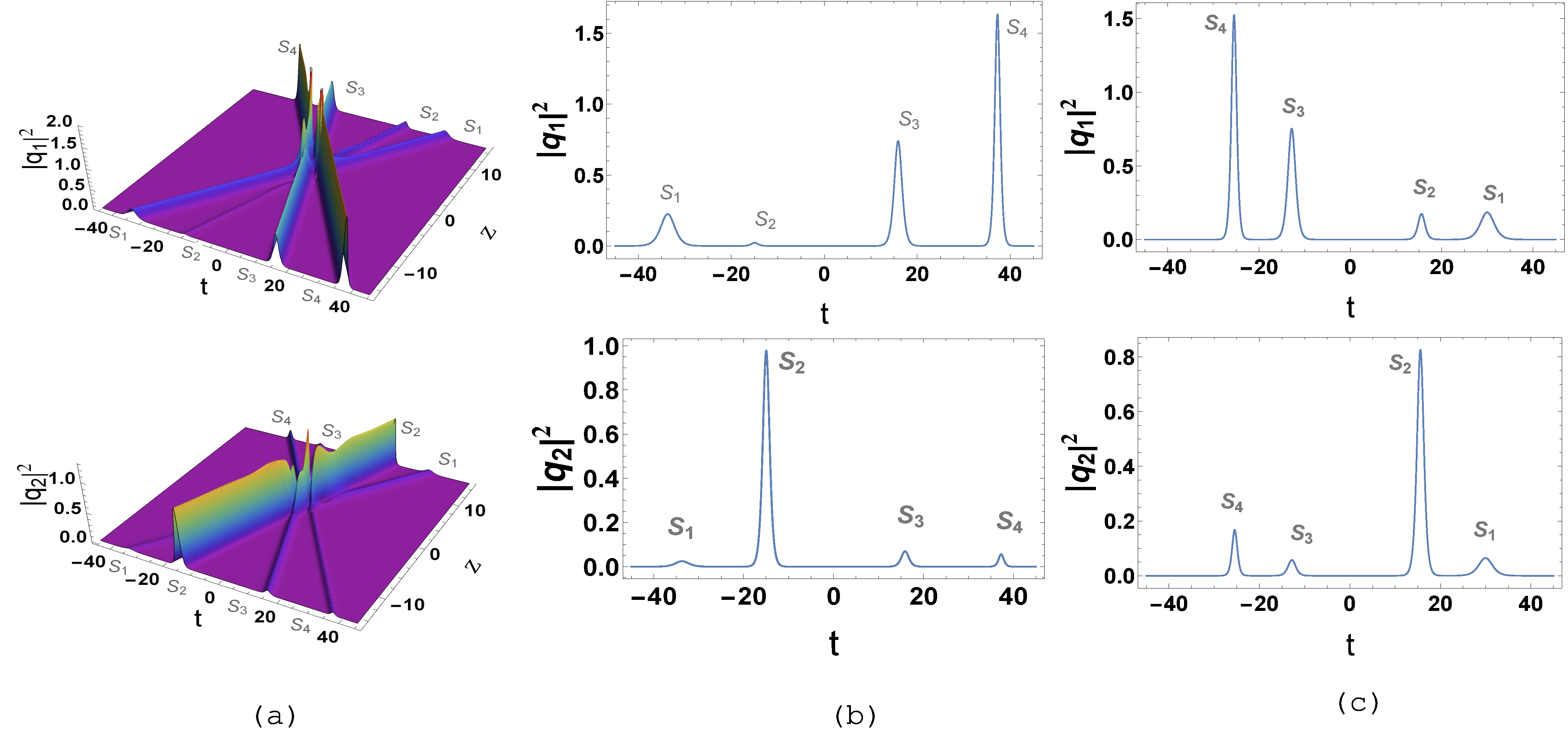}
\caption{OR gate: Input states  $``1"$ and  $``0"$; Output state  $``1"$. }\label{or_10}
\end{center}
%\end{figure}
%\begin{figure}
\begin{center}
\includegraphics[width=1\linewidth]{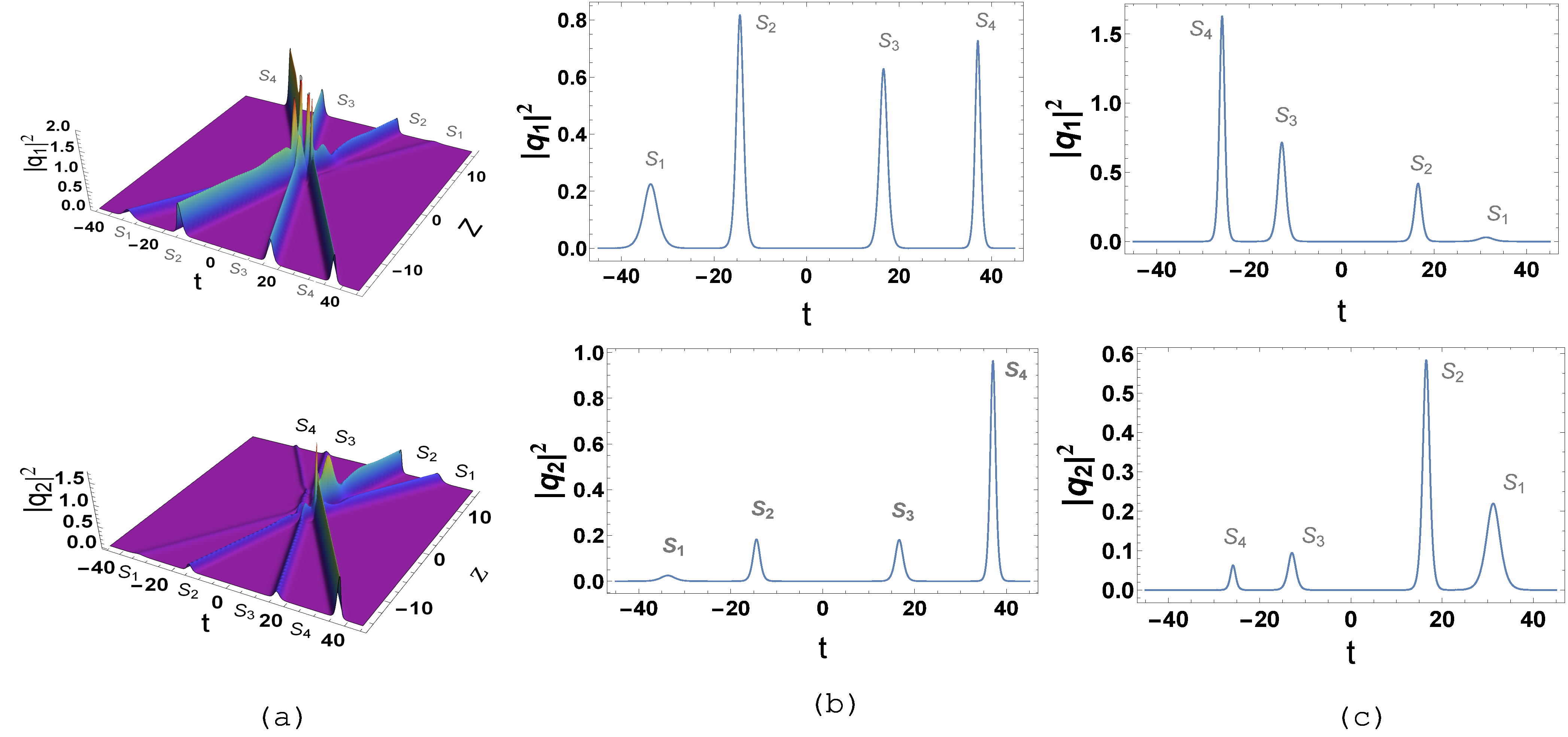}
\caption{OR gate: Input states  $``1"$ and  $``1"$; Output state  $``1"$. }\label{or_11}
\end{center}
\end{figure}
\begin{table}[!ht]
\begin{minipage}{.5\linewidth}
%\label{table1}
%{\scriptsizebre
\caption{Truth table of OR gate}
\centering
\begin{tabular}{|c|c|c|c|}
 % \cline{1-6}
\hline
 Input 1 ($S_1$) & Input 2 ($S_2$) & Output ($S_{4}'''$) \\\hline
  0& 0 & 0  \\ \hline
   0& 1& 1  \\ \hline
1& 0 & 1  \\ \hline
1& 1 & 1  \\ \hline
\end{tabular}
\end{minipage}%
\begin{minipage}{.5\linewidth}
%\scriptsize
\caption{Intensity table of OR gate}
\begin{tabular}{|c|c|c|c|}
 % \cline{1-6}
\hline
Input 1 ($S_1$) (W)  & Input 2 ($S_2$)(W)  & Output ($S_{4}'''$) (W)  \\\cline{1-3}
$|\rho^{1-}|^2$ &$|\rho^{2-}|^2$ &$|\rho^{4+}|^2$ \\ \hline
  0.1& 0.3 & 0.7  \\ \hline
   0.1& 32& 40  \\ \hline
10& 0.02 & 8  \\ \hline
9& 5 & 27  \\ \hline
\end{tabular}
\end{minipage}
\end{table}